\documentclass[letterpaper,twocolumn,showpacs,preprintnumbers,amsmath,amssymb]{revtex4}

\usepackage{graphicx}
\usepackage{dcolumn}
\usepackage{bm}
\usepackage{array}

\begin{document}

\preprint{Accepted for publication in {\em Phys. Rev. A.}}

\title{Resonant Electric Dipole-Dipole
Interactions between Cold Rydberg Atoms in a Magnetic Field}%

\author{K.~Afrousheh}
\author{P.~Bohlouli-Zanjani}
\author{J.~D.~Carter}
\author{A.~Mugford}
\author{J.~D.~D.~Martin}
\affiliation{%
Department of Physics and Institute for Quantum Computing \\
University of Waterloo, Waterloo, ON, N2L 3G1, CANADA
}%

\date{April 21, 2006}

\begin{abstract}
Laser cooled $^{85}{\rm Rb}$ atoms
were optically excited to $46d_{5/2}$ Rydberg states.  A microwave
pulse transferred a fraction of the atoms to the $47p_{3/2}$ Rydberg
state.  The resonant electric dipole-dipole interactions between
atoms in these two states were probed using the linewidth of the
two-photon microwave transition $46d_{5/2}-47d_{5/2}$.  
The presence of a weak magnetic field $\approx 0.5 \: {\rm G}$ 
reduced the observed line broadening, 
indicating that the interaction is suppressed by
the field.  The field removes some of the energy degeneracies
responsible for the resonant interaction, and this is the basis
for a quantitative model of the resulting suppression.
A technique for the calibration of magnetic field strengths using
the $34s_{1/2}-34p_{1/2}$ one-photon transition is also presented.

\end{abstract}


\maketitle

\section{Introduction}

The interactions between neighboring gas phase atoms 
can be enhanced
by exciting them to Rydberg states.  
To obtain translationally cold interacting Rydberg gases 
($\approx 100 \: \mu K$),
laser cooled atoms can be excited \cite{anderson:1998,mourachko:1998}. 
The behavior of these cold Rydberg gases is 
often qualitatively different from that of hotter samples 
($\approx 300 \: {\rm K}$).  
For example, a cold cloud of Rydberg atoms can spontaneously evolve 
into a plasma under conditions in which a hot cloud could 
not \cite{li:2004}.  Recent theoretical 
calculations suggest that cold interacting Rydberg atoms on uniformly 
spaced lattices will show properties analogous to that of a plasmon
wave propagating along metal nanoparticles
\cite{robicheaux:2004}.

Translationally cold Rydberg atoms have recently
received attention in the context of neutral atom quantum information
processing.    Theoretical work has suggested that Rydberg atom
interactions may be used to implement gates between atoms holding
qubits \cite{jaksch:2000},  for encoding qubits in multiatom 
clouds \cite{lukin:2001}, and to implement
``on demand'' single-photon sources \cite{saffman:2002}.  
Recent experimental work has demonstrated a suppression of the average
Rydberg density due to interactions -- so called ``local blockade''
\cite{tong:2004,singer:2004},  and rapid progress is expected in this 
area over the next several years \cite{ryabtsev:2005}. 

Transitions between Rydberg states can be driven using
microwave sources, which offer stability, linewidth, and modulation 
capabilities unrivaled by lasers.  
To elucidate the role of multibody interactions in
cold Rydberg gases, Mourachko
{\it et al.}~\cite{mourachko:2004}
used microwave radiation
to transfer a fraction of the optically excited
Rydberg atoms into a different angular momentum state. 
Recently, Li {\it et al.}~\cite{li:2005}
have used a similar microwave redistribution
pulse to establish the role of the electric dipole-dipole
interaction in plasma formation from cold Rydberg gases.
In addition,  the linewidths of microwave driven transitions between 
Rydberg states can be used as a sensitive probe of interactions 
between Rydberg atoms \cite{afrousheh:2004}.

Recently, we have probed
the strong resonant electric dipole-dipole interactions between
$^{85}{\rm Rb}$ atoms in $45d_{5/2}$ and $46p_{3/2}$ states
using the linewidth of a
two-photon microwave transition from $45d_{5/2}$ to $46d_{5/2}$
\cite{afrousheh:2004}.
The results were consistent
with theoretical expectations, and 
this approach promises 
to be a powerful technique for 
studying cold Rydberg atom interactions.

How will weak magnetic fields ($\approx 0.5 \: {\rm G}$) 
influence the interactions between Rydberg atoms?
Some of the energy degeneracies which are responsible 
for the resonant electric dipole-dipole interaction 
observed in Ref.~\cite{afrousheh:2004} can be spoiled 
by application of a magnetic field.  
It may be desirable to excite neutral atoms to Rydberg states
in a Ioffe-Pritchard trap without switching off
the trapping fields.  Rydberg atoms themselves could also
be trapped using inhomogeneous magnetic fields \cite{lesanovsky:2005}.  

In the present
work we use microwave spectroscopy of Rydberg atoms to calibrate
DC magnetic field strengths.
Microwave fields are then used to manipulate Rydberg populations,
and probe interactions, demonstrating that
the dipole-dipole interaction between cold Rydberg atoms is 
partially  suppressed by magnetic fields. A quantitative model is presented 
which explains these results.

\section{Experimental}

The apparatus is similar to that described in Ref.~\cite{afrousheh:2004}.
Cold $^{85}$Rb atoms in a vapor-cell 
magneto-optical trap (MOT) \cite{monroe:1990}
are excited to high-$n$ Rydberg states using a pulse of laser
light.  Microwave pulses are then used to transfer Rydberg population
and probe interactions.  The final state populations are analyzed by
selective field ionization (SFI) \cite{gallagher:1994}.  
The significant differences
between the present apparatus and that in Ref.~\cite{afrousheh:2004}
are the laser excitation source and
the control of the magnetic field during the experiment.

Instead of a pulsed dye laser system, a frequency-doubled cw 
Ti:sapphire system is used to excite Rydberg states.
A commercial Ti:sapphire ring laser (Coherent MBR-110) operating
at approximately 960 nm is frequency doubled
in an external ring resonator (Coherent MBD-200) 
to produce approximately
90 mW at 480 nm.    Light pulses of $1 \: \mu s$
duration are produced using an acousto-optic modulator.
After the acousto-optic modulator the beam is coupled into a single
mode fiber.  The output beam from the fiber is collimated then focused
and directed towards the trapped atoms.  
The highest Rydberg densities 
($\approx 2 \times 10^7 \: {\rm cm}^{-3}$) 
in this paper were obtained with optical 
powers of approximately $25 \: {\rm mW}$.  

Excitation of cold atoms to Rydberg states occurs 
as a two-color, two-photon process with the 480 nm light, and
the nearly resonant, red detuned 780 nm light used for cooling and 
trapping \cite{teo:2003}.
Unfortunately, there are several atmospheric 
water absorption lines in the vicinity of 960 nm which restrict
performance of the Ti:sapphire laser system.  To avoid one of
these, we excite $46d_{5/2}$ states instead of the
$45d_{5/2}$ states used in Ref.~\cite{afrousheh:2004}. 

Microwave horns are aimed at the Rydberg atoms through a fused silica window.  
To drive the 
$46d_{5/2} - 47d_{5/2}$ ($\approx 35.7 \: {\rm GHz}$), 
the $46d_{5/2} - 47p_{3/2}$ ($\approx 22.1 \: {\rm GHz}$), and 
the $46d_{5/2} - 45f_{5/2}$ ($\approx 24.1 \: {\rm GHz}$)
transitions we directly use the unamplified output of 
Agilent E8254A microwave synthesizers.  
For the 
$46d_{5/2} - 46p_{3/2}$ ($\approx 100.1 \: {\rm GHz}$) 
and the 
$34s_{1/2} - 34p_{1/2}$ ($\approx 104.1 \: {\rm GHz}$) 
transitions both 
an active quadrupler (Spacek Labs P/N A100-4XW) or 
a passive tripler (Pacific Millimeter P/N W3WO) 
were used.  These multipliers were driven by an 
Agilent E8254A synthesizer.

Two non-magnetic stainless steel electrode plates, separated by 
$36 \:{\rm mm}$, are located to either side of the trapped atoms.
These electrodes are used for SFI
of Rydberg atoms \cite{gallagher:1994}.
The resultant ions are drawn by the same fields 
through a 6 mm hole in one of the electrode plates
towards a microchannel plate 
detector. 
These plates are also useful for controlling the electric field
in the experimental region.  By varying the relative voltage difference 
of the two plates, their average voltage and the voltage of the 
Rb dispenser
source with respect to the grounded chamber, it is possible to compensate 
for stray electric fields in all three directions in the experimental 
region.  Stray electric fields are measured using the Stark shifts
of a one-photon microwave transition \cite{osterwalder:1999}.
The residual electric field is estimated to be less than 0.1 V/cm.

To obtain estimates of the Rydberg density it is necessary
to know the number of Rydberg atoms and their spatial extent.
The spatial profile of the 480 nm beam determines the extent
of the Rydberg cloud in two directions (the focus of this beam
is much smaller than the trapped atom dimensions).
Beam characterization cannot easily be done within the vacuum chamber.
Instead, the rigidly attached output coupler and focusing assembly are 
moved to an optical table and scanning knife edge 
beam profile measurements are 
performed throughout the focal region
(see for example Ref.~\cite{suzaki:1975}).  The beam profile is
Gaussian.
Since there are collision processes which depend on the Rydberg density,
the location of the tightest 
focus in the trap can be determined by moving 
the focusing lens relative to the trap to maximize these processes.  
Once this location is found, the excitation
beam profile can be systematically varied by moving the 
focusing lens.
The beam was set to have a 
full width at half maximum (FWHM) of $0.190\pm 0.015 \: {\rm mm}$ at the
location of the trap.

The extent of the Rydberg cloud in the third dimension is
determined by the trapped atom profile.
By scanning the direction of the $480 \: {\rm nm}$ excitation laser --
moving the beam over the trapped atom cloud --
it is observed that there
is an excellent correlation between the Rydberg signal 
and $780 \: {\rm nm}$
fluoresence imaging measurements of cloud extent 
(accounting for the finite width of the $480 \: {\rm nm}$ beam).
Therefore $780 \: {\rm nm}$ 
fluorescence imaging gives the spatial extent of 
the Rydberg cloud in the direction of the $480 \: {\rm nm}$ 
beam propagation.  This is typically $0.5 \pm 0.1$  mm (FWHM).

The beam profile of the 480 nm Rydberg excitation laser and 
the spatial profile of the trapped atoms are combined to 
convert the observed field ionization signals into 
spatially averaged Rydberg densities.  
If the waist of the excitation laser is significantly smaller than
the width of the trap, and this waist does not change appreciably 
as the beam passes through the trap
(conditions that are approximately satisfied in the present setup) 
the spatially averaged Rydberg density corresponding
to a total number of Rydberg atoms $N_{Ryd}$ is
\begin{equation}
n_{Ryd} = \left( \frac{2 \ln 2}{\pi}  \right)^{3/2} \:
\frac{N_{Ryd}}{({\rm FWHM}_{\perp})^2 ({\rm FWHM}_{\parallel})}
\label{eq:dens}
\end{equation}
where ${\rm FWHM}_{\perp}$ 
is the full width at half maximum of the irradiance profile of the
480 nm light, and ${\rm FWHM}_{\parallel}$ corresponds to the 
$5p_{3/2}$ spatial distribution in the direction of the 480 nm
beam.

Determination of $N_{Ryd}$
requires knowledge of the absolute detection efficiency and
gain of the microchannel plate detector (MCP).
The absolute detection efficiency of an MCP
depends on the kinetic energy at impact, the species involved,
and surface conditions \cite{fraser:2002}.
A recommended efficiency for our specific conditions was not available.
Based on information from the manufacturer and 
the literature \cite{fraser:2002}
we estimate an initial detection efficiency of 50\% -- however
this could be as high as 80\% or as low as 20\%.
The gain of the MCP was calibrated by observing
the average total charge due to single events at higher 
bias voltages than normal operating conditions.  
To determine gain at the operating bias voltage,
a series of measurements were taken at constant charged particle 
signal level with variable bias voltages.

The average densities quoted in this paper are obtained using 
Eq. \ref{eq:dens}.
The uncertainty in these is dominated by $N_{Ryd}$,
which is limited by the initial detection efficiency estimate.

\begin{figure}
\includegraphics{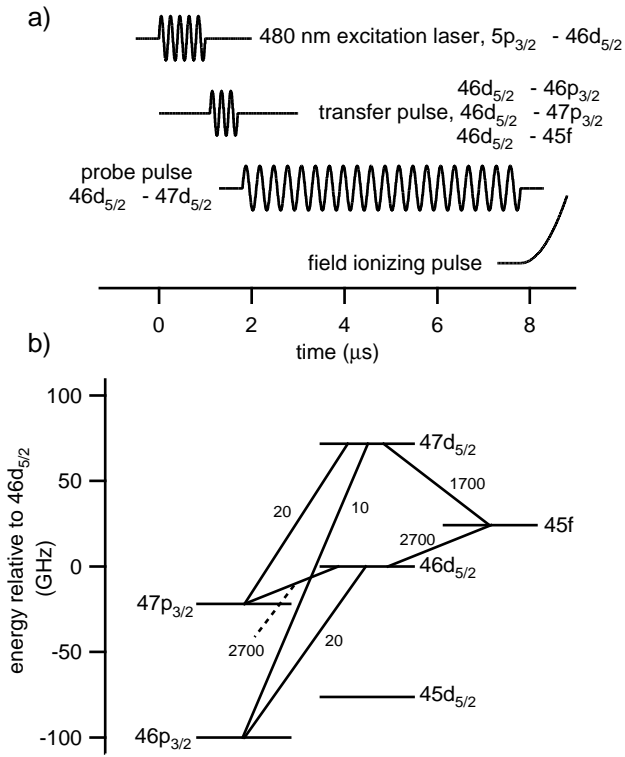}
\caption{\label{fg:timing}
a) Timing for experiment.  There are delays of roughly 100 ns between
the excitation laser pulse, transfer pulse, probe pulse and
field ionizing pulse.  This sequence is repeated at 
$10 \: {\rm Hz}$ and is synchronized with the 
$60 \: {\rm Hz}$ AC power line.
b) 
Energy levels of relevant states.
The spectroscopic data is from Ref.~\cite{li:2003}.  
The matrix elements magnitudes
$|\!\!<\!\!n\ell_j|r|n'\ell'_{j'}\!\!>\!\!|$
are shown between appropriate states, where $r$ is
the radial coordinate of the Rydberg electron.
These are expressed in atomic units, and were calculated using
the techniques described by 
Zimmerman {\it et al.}~\cite{zimmerman:1979}.
}
\end{figure}

\section{Observation of the Resonant Electric Dipole-Dipole
Interaction}
\label{se:observationbasic}

The interactions between $^{85}{\rm Rb}$
Rydberg atoms were studied using the
linewidths of the $46d_{5/2} - 47d_{5/2}$ 
two-photon ``probe'' transition.
Li {\it et al.}~\cite{li:2003}
have shown that two-photon transitions between
states with identical $g_J$ factors, such as $46d_{5/2}$
and $47d_{5/2}$, show negligible broadening in the
inhomogeneous magnetic field of a MOT.
These transitions also have less sensitivity to electric fields
than one-photon transitions.

The typical experimental sequence for a single shot is
shown in Fig.~\ref{fg:timing}a).  
Depending on the particular
situation, a ``transfer pulse'' may be used shortly after optical 
excitation to drive a fraction of the initially excited $46d_{5/2}$ atoms
into other states.  In the presence of the inhomogeneous magnetic field of
the MOT any coherent superpositions formed by this pulse will rapidly
dephase across the sample.  Interatomic interactions are somewhat
inhomogeneous due to the variation in spacings between atoms, and
are also expected to cause appreciable dephasing.  
As a consequence the transfer pulse is viewed as redistributing
population rather than creating superpositions.

The $46d_{5/2} - 47d_{5/2}$ probe microwave frequency
is scanned between shots.  The power of this probe 
is adjusted so that power broadening is negligible, and in
the limit of low Rydberg densities, the linewidths of this transition
are entirely dominated by transform broadening.
To analyze additional broadening due to
interatomic effects, the spectra are fit to 
the transform limited lineshape for a square excitation pulse
(${\rm sinc}^2(\pi f T)$)
convolved with a Lorentzian of variable width $\delta \nu$.  
The widths of the fitted Lorentzian are then analyzed as 
a function of spatially averaged total Rydberg density 
(which is normally adjusted using the $480 \: {\rm nm}$ excitation
laser power).

\begin{figure}
\includegraphics{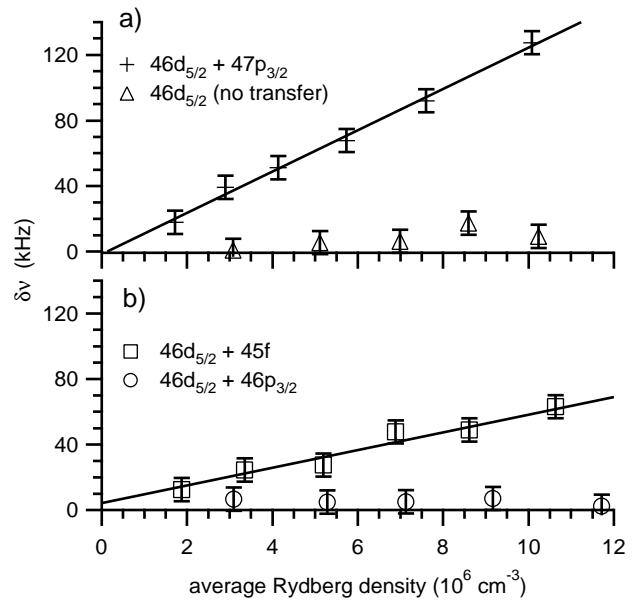}
\caption{\label{fg:linewidth}
Broadening of the $46d_{5/2}-47d_{5/2}$ probe transition as a function
of average Rydberg density with and without transfer pulses.
The transfer pulses create 50 \% mixtures, except in the case of
the $46d_{5/2}+45f$ mixture, where only 30 \% of the total Rydberg
atoms are transferred to the $45f$ state.  The straight lines are
least squares fits.  See the text for the definition
of $\delta \nu$.
}
\end{figure}

As illustrated in Fig.~\ref{fg:linewidth}, $\delta \nu$ converges
to zero as the Rydberg density goes to zero.
This indicates that the line broadening is an interatomic effect.
At $1 \times 10^{7} \: {\rm cm^{-3}}$ Rydberg density,
transferring one-half of the $46d_{5/2}$ atoms to $47p_{3/2}$ 
broadens the linewidth, increasing
$\delta \nu$ from approximately $10 \: {\rm kHz}$ to 
$120 \: {\rm kHz}$ (the uncertainty in $\delta \nu$ is
typically $\pm 7 \: {\rm kHz}$).  This can be attributed to
the resonant electric dipole-dipole interaction between atoms
in these two states.  Consider a pair of atoms labelled A and B.
In the absence of interactions between these atoms, the state 
$|46d_{5/2}m_{5/2}\!\!>_{\rm A}\!\! |47p_{3/2}m_{3/2}\!\!>_{\rm B}$ 
is energy degenerate with 
$|47p_{3/2}m_{3/2}\!\!>_{\rm A}\!\!|46d_{5/2}m_{5/2}\!\!>_{\rm B}$.
The dipole-dipole interaction breaks this degeneracy, leading to
an energy splitting on the order of  
\begin{equation}
\Delta \nu_{dd} \approx 
\frac{|\!<46d_{5/2}|r|47p_{3/2}>\!|^{2}}{R_{\rm AB}^3}
\label{eq:rough}
\end{equation}
where $r$ is the separation of Rydberg electron from the ion-core,
and $R_{\rm AB}$ is the typical interatomic spacing 
\cite{fioretti:1999,anderson:2002}.

The value of $|\!\!<46d_{5/2}|r|47p_{3/2}>\!\!|$ and other relevant radial 
transition dipole moments are indicated in Fig.~\ref{fg:timing}b).
For instance, the dipole coupling between the $47d_{5/2}$
and $47p_{3/2}$ states is relatively weak compared to the coupling
between the $46d_{5/2}$ and $47p_{3/2}$ states.
Thus the broadening of the two-photon $46d_{5/2} - 47d_{5/2}$ probe 
transition beyond the transform limit is 
dominated by dipole-dipole energy splittings in the initial state.  
As in Ref.~\cite{afrousheh:2004}, a simple estimate
based on Eq.~\ref{eq:rough} is consistent with the observed broadening.  
The $1/R_{\rm AB}^3$ scaling of the splittings implies a linear density
dependence of the line broadening, which is clearly demonstrated in
Fig.~\ref{fg:linewidth}a).

The dipole-dipole coupling between $46d_{5/2}$ and $45f$ states is
of comparable magnitude to that between $46d_{5/2}$ and
$47p_{3/2}$ states (see Fig.~\ref{fg:timing}b).
Consequently, transfer of a fraction of atoms to the $45f$ state
gives similar broadening (see Fig.~\ref{fg:linewidth}b).

It is also straightforward to drive transitions to a nearby state
which is not strongly dipole coupled to either the initial or final
state of the two-photon probe transition.  
Based on the radial dipole matrix elements we would expect
4 orders of magnitude less broadening if 50\% of the atoms were
transferred to $46p_{3/2}$ instead of $47p_{3/2}$.
As Fig.~\ref{fg:linewidth}b) indicates, there is no detectable
broadening of the probe transition in the presence of $46p_{3/2}$
atoms.  

These results are similar to those presented in Ref.~\cite{afrousheh:2004}
(only at one higher principal quantum number).  There is a slight
increase in linewidth of $\approx 20 \% $ in a mixture of 
$nd_{5/2}$ and $(n+1)p_{3/2}$ atoms (where $n=46$ here, and $n=45$
in Ref.~\cite{afrousheh:2004}).
The energy splittings $\Delta \nu_{dd}$ scale like $n^4$ due to the
$n^2$ scaling of the transition dipole moments \cite{gallagher:1994},
and this accounts for about a 10 \% increase.  We have also made
changes in our Rydberg density estimate procedure, and these may
also account for the discrepency.
The more significant difference is in the linewidths obtained when
no transfer pulse is present (see Fig.~\ref{fg:linewidth}).
There is roughly 3 times less broadening as compared to
the data presented in Ref.~\cite{afrousheh:2004}.

The source of broadening in the case of no transfer pulse was proposed
to be due to the collision 
of hot Rydberg atoms ($\approx 300 \: {\rm K}$)
with the cold Rydberg atoms \cite{afrousheh:2004}.  
These types of collisions have also been 
discussed in the context of cold plasma formation \cite{robinson:2000}.
The hot atoms are excited in the same way as the cold trapped atoms.
They make an undetectable contribution to the observed 
field ionization  because the waiting period between excitation
and extraction allows them to move to where they can't be detected.

Several experimental tests support the role of the hot Rydberg atoms
in line broadening.  To control the hot background Rb number density, 
the current through an alkali metal dispenser
source is varied \cite{fortagh:1998}.
When the hot Rb background density is varied in this manner,
the 480 nm excitation laser power should be adjusted to produce the
same cold Rydberg density (compensating for the different trapped atom 
density).  The time constant is relatively long
for changes in background pressure with dispenser current 
(on the order of 10 minutes), so we can check
independently of background Rb pressure
whether thermal radiation or charged particles from the dispenser
are responsible for broadening.
At our sensitivity, no such effects were
found.

With constant cold Rydberg density, it is found that $\delta \nu$ 
increases with hot Rb background density.
This rules out the interactions between cold Rydberg atoms as 
being responsible for the broadening.  
However, at higher densities than those studied
here, the nearly resonant interactions 
$46d_{5/2}46d_{5/2} \leftrightarrow 48p_{3/2}44f$ 
between cold Rydberg atoms will become increasingly important \cite{li:2005}.
At a fixed hot Rb
background density, $\delta \nu$ is proportional to 480 nm
excitation laser power, which rules out the influence of unexcited
hot atoms.

Not only does the hot Rydberg atom density increase with 
background Rb pressure, but so does the trapped atom density.
This suggests a check to determine whether interactions with less
excited cold atoms, or with hot Rydberg atoms from the vapor, 
are the cause of the increased broadening.
If the cooling and trapping laser frequency is detuned
to obtain less cold atoms, and the 480 nm laser power is increased
to obtain the same cold Rydberg density, the observed linewidths
increase, because more hot atoms are optically excited to Rydberg
states.  If less excited cold atoms were responsible for the
broadening we would expect a decrease in linewidth.

In summary, the broadening $\delta \nu$ observed with no transfer pulse
is due to the collision of hot Rydberg atoms with the cold Rydberg
atoms we are studying.
In the present work $\delta \nu$ in the absence of a transfer pulse is
smaller than in Ref. \cite{afrousheh:2004} for at least two
reasons.  We are now working at lower dispenser currents
(corresponding to lower background pressures) and using a 
$480 \: {\rm nm}$ excitation source with significantly narrower bandwidth.
The reduced bandwidth is expected to alter the ratio of cold to hot
Rydberg atoms.  

By working with a low hot Rb background pressure, and at
$10^7 \: {\rm cm}^{-3}$ total Rydberg density, $\delta \nu$ is
on the order of its experimental uncertainty $\pm 7 \: {\rm kHz}$.
We have not yet established what the ultimate limit on the 
lowest $\delta \nu$ is,  and thus the ultimate sensitivity of the 
probe transition to Rydberg interactions.  However, the currently 
demonstrated sensitivity is vastly superior to what can readily
be achieved with optical probes.

\section{Magnetic Field Measurements}

In this section we discuss the use of a 
single-photon microwave transition between Rydberg states
to measure magnetic fields.

\begin{figure}
\includegraphics{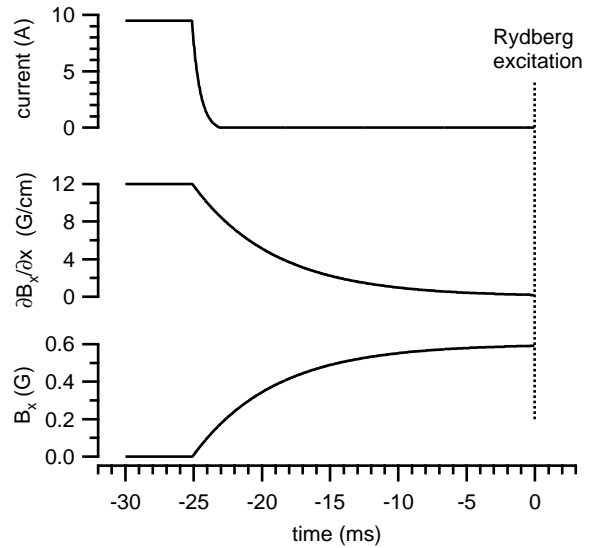}
\caption{\label{fg:magtiming}
Timing for experiments done in the presence of a homogeneous
DC magnetic field.
The quadrupole field coil current
is cut to zero in $1.7 \: {\rm ms}$.  As the magnetic field
gradient $\partial B_x / \partial x$ decays to zero, 
the field in the center of the
trap $B_x$ gradually increases due to the presence of a non-zero
homogeneous bias field.  At $t=0$,
excitation to Rydberg states and subsequent microwave probing
takes place, as shown in Fig.~\ref{fg:timing}.  
The gradient and bias field time constant was measured 
using Rydberg atom microwave spectroscopy (see text).
}
\end{figure}

The MOT used to trap the $^{85}$Rb
atoms requires a magnetic quadrupole field \cite{monroe:1990}.  
In these experiments
the field gradients in the three orthogonal directions are 
estimated to be 12, 12 and 24 G/cm, based on the coil geometry.
The Rydberg cloud has a Gaussian profile with FWHM's of 
$0.190 \pm 0.015 \: {\rm mm}$ in two dimensions
(dictated by the 480 nm excitation laser)
and $0.5 \pm 0.1 \: {\rm mm}$ in the third dimension
(dictated by the trap size).
The largest dimension corresponds to the direction of
the largest magnetic field gradient.  This gives a root mean
square (RMS)
magnetic field of approximately $0.5 \: {\rm G}$.  
Since Zeeman shifts
are typically on the order of 1 MHz/G, and the dipole-dipole
interactions that we observe are roughly 100 kHz, it is
apparent that the quadrupole field is a significant
inhomogeneity, and it is desirable to turn this off.
The current in the coils producing 
this field is shut off in $1.7 \: {\rm ms}$, 
but significant eddy currents
persist for much longer (see, for example, Ref.~\cite{dedman:2001}).
Excitation to Rydberg states occurs after a delay of 
$25 \: {\rm ms}$, to
allow the eddy currents to die down 
{\bf (see Fig.~\ref{fg:magtiming})}.
As will be discussed, an upper bound can be placed
on any remaining inhomogeneity.

The detuning and counterpropagating beam balance of the MOT
lasers are optimized to minimize the reduction in cold atom
density during the  $25 \: {\rm ms}$ delay.  
Immediately after selective field ionization is complete, the current
to the quadrupole coils is turned back on to recapture cold
atoms for the next shot.

Once the quadrupole field is turned off,  a DC magnetic field may remain.
Three pairs of orthogonal Helmholtz coils are used to null
any preexisting field (due to the ion-pump, earth etc...)
and one of these coil pairs is used to produce a deliberate 
homogeneous field.  The current in the coils is constant throughout
the MOT collection, Rydberg excitation, and microwave probe cycle.
When setting a non-zero homogeneous field, the quadrupole
zero of the MOT will be shifted.  As long as the quadrupole zero
remains significantly within the overlap of the cooling 
and trapping beams, the trap will still function.  
The use of fields of up to 0.6 G pointing in the direction with a field
gradient of 12 G/cm, shifts the zero by 0.5 mm, which does not 
significantly influence MOT operation when using 
780 nm beams with a FWHM of $4 \: {\rm mm}$.
Loss of the cold atoms in the $25 \: {\rm ms}$
waiting period limits the DC bias fields which can be applied using 
this technique.  

Although it is straightforward to calculate the fields due to
the bias coils from their geometries, we do not {\it a priori}
know the stray magnetic field within the apparatus.
To determine this, the one-photon microwave transition
$34s_{1/2}-34p_{1/2}$ of $^{85}$Rb is used.
We work at a lower $n$ 
compared to where the dipole-dipole experiment is performed
so that the hyperfine spacings are larger (these scale
like $1/n^3$, see Ref.~\cite{gallagher:1994}).
To improve the sensitivity of the measurement it is desirable
to use long microwave pulse lengths (for narrow linewidths).  
Redistribution of Rydberg state population due to thermal radiation, 
limits the microwave pulse length.
Pulses of $36 \: \rm {\mu s}$ are used.

\begin{figure}
\includegraphics{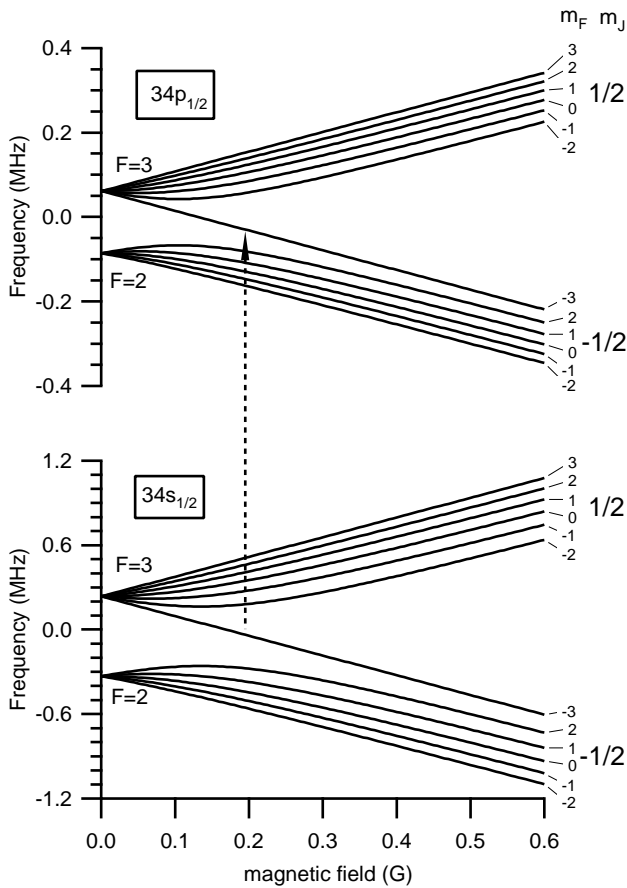}
\caption{\label{fg:oneelevels}
Calculated Zeeman structure of the $34s_{1/2}$ and $34p_{1/2}$
levels of $^{85}{\rm Rb}$ using the Breit-Rabi 
formulae \cite{kuhn:1962} (the $m_J$ labelling is approximate).
Please note the different
vertical scalings.  The two hyperfine splittings are chosen to
match the spectral lines in Fig.~\ref{fg:onespectra}
($0.567 \: {\rm MHz}$ for $34s_{1/2}$, and 
$0.147 \: {\rm MHz}$ for $34p_{1/2}$).
The arrow labels the 
$34s_{1/2}(m_F\!=\!-3)$ to 
$34p_{1/2}(m_F\!=\!-3)$ 
line, which is used for magnetic field calibration (see text).
}
\end{figure}

\begin{figure}
\includegraphics{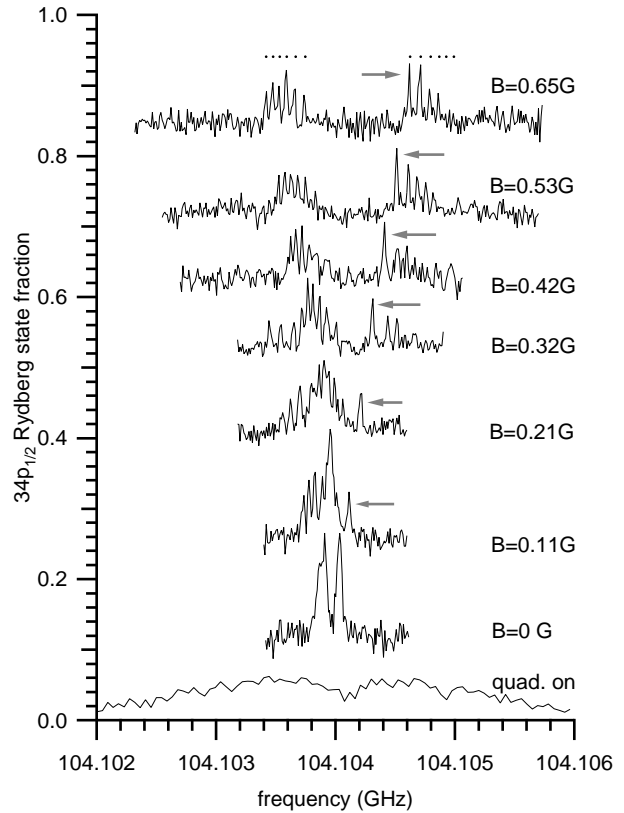}
\caption{\label{fg:onespectra}
The $^{85}{\rm Rb}$ $34s_{1/2}$ to $34p_{1/2}$ microwave spectra taken
with different DC magnetic
fields (top 7 traces)
and with the MOT quadrupole field present (bottom trace).
Spectra are offset vertically for clarity.  The calculated line
positions of the $\Delta m_F=0$ transitions using the energy levels
from Fig.~\ref{fg:oneelevels} are shown for the 
$0.65 \: {\rm G}$ data (the microwave polarization vector is in same
direction as the applied magnetic field).
The arrow labels the 
$34s_{1/2}(m_F\!=\!-3)$ to 
$34p_{1/2}(m_F\!=\!-3)$ 
line, which is used for magnetic field calibration (see text).
}
\end{figure}

The Zeeman splittings of the relevant levels are shown in 
Fig.~\ref{fg:oneelevels}.
The 780 nm light used for optical excitation is nearly
resonant with only the $5s_{1/2}F=3$ to $5p_{3/2}F=4$ transition.
Since $F$ cannot change by more than one in a single photon transition,
the $34s_{1/2}F=2$ state is not excited in zero magnetic field,
and only two microwave spectral lines are observable: 
$34s_{1/2},F=3$ to $34p_{1/2},F=2$ and 
$34s_{1/2},F=3$ to $34p_{1/2},F=3$
(see Fig.~\ref{fg:onespectra}).

As a magnetic field is applied, this spectrum splits into many lines.
To zero the magnetic field in a specific direction
the following procedure is followed.  
A current  is applied to a set 
of coils to produce a relatively large magnetic field (the precise
value is unknown at this point).  By scanning the microwave frequency,
a spectrum with many lines can be observed 
(see Fig.~\ref{fg:onespectra}).
Then we determine the current which produces the same field magnitude
{\em in the opposite direction}, 
by trying to reproduce the spectral
line positions.  The average of these two currents corresponds
to zero field projection along this axis.  
Using this procedure all three orthogonal axes can be zeroed
independently.  Since the {\em change} in magnetic field with current
can be estimated from coil geometry, the precision to which the
average currents can be determined can be directly related to the
precision to which the fields can be compensated.  
We have zeroed the magnetic field along
each axis to within $\pm 6 \: {\rm mG}$
using this technique.

To deliberately create magnetic fields in a specific direction,
one of the coil currents was adjusted
away from its ``zero'' value.
The change in magnetic field
with coil current can be estimated from the coil geometry.
However, the line positions in the $34s_{1/2}-34p_{1/2}$ spectra 
(see Fig.~\ref{fg:onespectra}) can be used to directly measure the field.
The shifting of energies is given by the 
Breit-Rabi formulae (see, for example Ref.~\cite{kuhn:1962}).
In general, line positions are dependent on the magnetic
field, and the hyperfine splitting of both the $34s_{1/2}$
and $34p_{1/2}$ states.  The $34s_{1/2}$ hyperfine spacing
is not known {\it a priori}.  However, there is a 
particular line (marked with arrows in Fig.~\ref{fg:onespectra}),
corresponding to the transition from 
$34s_{1/2}(m_F\!=\!-3)$ to 
$34p_{1/2}(m_F\!=\!-3)$ 
whose shift in position with magnetic field
does not depend on the hyperfine spacing.
The transition energy for this line is
$(2/3) \mu_{B} B $, where $\mu_{B}$ is the Bohr
magneton and $B$ is the magnetic field.
As apparent in Fig.~\ref{fg:onespectra} this line is convenient as
it is readily identifable in the spectra.
By tabulating the position of this spectral line as a function
of coil current and fitting a straight line, the magnetic field
as a function of coil current can be determined, without knowledge
of the $34s_{1/2}$ hyperfine splitting.
This provides a precise determination of the magnetic
field ($\pm 10 \:{\rm mG}$).
The dipole-dipole interaction is then studied under 
these calibrated magnetic field conditions.

It is desirable to know the inhomogeneity in the magnetic field 
when the experiment is performed. The bottom trace in 
Fig.~\ref{fg:onespectra} shows inhomogeneous broadening due to
the quadrupole field when the coil current is not shut off.
By varying the time between between shutting the current
off and excitation of the atoms to Rydberg states 
(see Fig. \ref{fg:magtiming})
we can observe the decay of this inhomogeneity using the spectral
widths.  The decay is observed to be exponential with a time
constant of $6 \: \pm 1 \: {\rm ms}$.  
A gaussmeter on the outside
of the vacuum chamber, placed as close as possible to the experimental
region (within $15 \: {\rm cm}$)
shows a similar time constant for the change in magnetic field:
$6.5 \: \pm 0.5 \: {\rm ms}$.

At the start of this section, the RMS magnetic field due to the 
quadrupole field was calculated to be 0.5 G.  Based on the exponential 
decay of the gradient, 25 ms after the coil currents are shut-off
we would expect this to decay to 8 mG.
This inhomogeneity is insignificant in the dipole-dipole interaction
experiments discussed in the next section.

The widths of the resolved 
$34s_{1/2}(m_F\!=\!-3)$ to 
$34p_{1/2}(m_F\!=\!-3)$ 
spectral lines in Fig.~\ref{fg:onespectra}
put an upper bound on the magnetic field inhomogeneity.  The lines
are observed to have a linewidth of $30 \: {\rm kHz}$.  
The transform limited linewidth for a 36 $\mu$s long pulse
is approximately $ 25 \: {\rm kHz}$.  If as a worst case
we assume that the linewidth contributions combine in quadrature,
the residual broadening due to magnetic field inhomogeneity and
other mechanisms is $ 17 \: {\rm kHz}$.  
This gives a rough upper bound
of 17 mG on the magnetic 
field inhomogeneity, which is consistent with the estimate
given in the previous paragraph.

\section{Suppression of the Resonant Electric
Dipole-Dipole Interaction Due to a Magnetic
Field}

Using the techniques described in the previous section, cold Rydberg atom
interactions can be studied in the presence of homogeneous, calibrated
magnetic fields.
The timing of the experiment remains the same as in 
Fig.~\ref{fg:timing}a).  However, the current to the quadrupole
coils is shut-off $25 \: {\rm ms}$ prior to photoexcitation
(see Fig.~\ref{fg:magtiming}).
During this waiting period, the cold atom cloud dimensions expand.
At each different applied magnetic field value, 
the counterpropagating beam powers
are re-optimized to minimize expansion.  When this is done, 
it is found that trap expansion is independent of DC magnetic field 
to within $10 \%$ for fields less than $0.6 \: {\rm G}$, and 
linewidth broadening can be studied at a fixed Rydberg density as a 
function of magnetic field.

\begin{figure}
\includegraphics{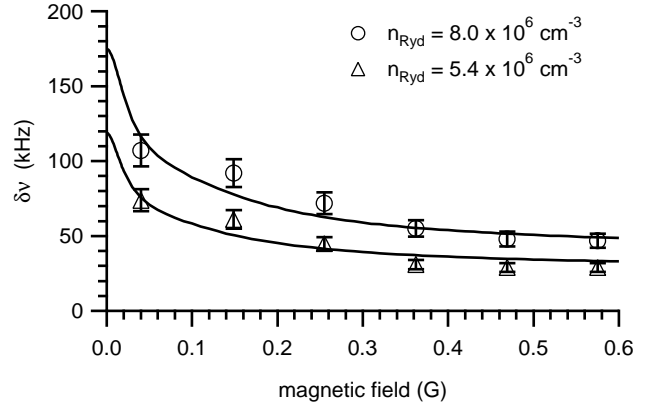}
\caption{\label{fg:suppression}
Broadening of the $46d_{5/2} - 47d_{5/2}$ probe transition 
in the presence of 50\% $47p_{3/2}$ atoms
as a function of magnetic field strength.
Two different total Rydberg number densities are shown.
The solid lines are calculations (see text for details).
}
\end{figure}

The linewidth broadening due to interatomic effects,
$\delta \nu$ is obtained in exactly the same manner as
described in Section \ref{se:observationbasic}.
In the limit of low densities $\delta \nu$ approaches zero for
all DC magnetic fields, indicating that it is due to interatomic
interactions. 
At an average Rydberg density of 
$8.0 \times 10^{6} \: {\rm cm^{-3}}$ we have found that
conversion of 50 \% of the $46d_{5/2}$ atoms to $47p_{3/2}$
increases 
$\delta \nu$ from 10 kHz to 110 kHz at approximately 0.04 G.
As Fig.~\ref{fg:suppression} indicates, increasing the
magnetic field reduces $\delta \nu$.  
The presence of a DC magnetic field
suppresses the resonant dipole-dipole interaction between atoms.
Essentially the magnetic field spoils some of the 
unperturbed atom energy degeneracies,
weakening the resonant interaction.

To discuss this in detail, we consider the dipole-dipole interaction 
operator:
\begin{equation}
\label{eq:vdd}
\hat{V}_{dd} = \frac{\vec{\mu}_{\rm A} \cdot \vec{\mu}_{\rm B}
    - 3 (\vec{\mu}_{\rm A} \cdot \vec{n}) 
(\vec{\mu}_{\rm B} \cdot \vec{n})}{R_{\rm AB}^3}
\end{equation}
where $\vec{\mu}_{\rm A}$ and 
$\vec{\mu}_{\rm B}$ are the electric dipole matrix element operators 
evaluated on each atom,  $\vec{n}$ is the unit vector pointing 
between the atoms, and $R_{\rm AB}$ is the separation of the two atoms.

The tensor product of the separated, non-interacting atom energy
eigenstates can be used as a basis set for the evaluation
of $\hat{V}_{dd}$. 
The axis of quantization for the projection of the
angular momentum of each atom
is chosen to point in the same direction 
as the magnetic field.  This is in contrast to the more conventional
choice for interatomic interactions, where the quantization
axis is chosen along $\vec{n}$ 
(see, for example, Ref.~\cite{santra:2003}).  This choice complicates
the evaluation of the matrix elements of $\hat{V}_{dd}$, but
simplifies the evaluation of the Zeeman part of the Hamiltonian.

If we label each atom as A and B,
in zero magnetic field with
no dipole-dipole interaction, all 24 states of the form 
$|1\!\!>=\!\!|46d_{5/2}m_{j,{\rm A1}}\!\!>_{\rm A}\!\!
|47p_{3/2}m_{j,{\rm B1}}\!\!>_{\rm B}$ 
are energy degenerate with the 24 states of the form 
$|2\!\!>=\!\!|47p_{3/2}m_{j,{\rm A2}}\!\!>_{\rm A}\!\!
|46d_{5/2}m_{j,{\rm B2}}\!\!>_{\rm B}$.
It is this degeneracy which is responsible for the 
{\em resonant} electric dipole-dipole line broadening.

Magnetic fields lift some of this 
degeneracy through the Zeeman effect.
If $| \psi_b \!\!>$ is a basis set vector
of the form given above, then:
\begin{equation}
\label{eq:zeeman}
<\!\! \psi_b|\hat{V}_{Z}|\psi_b \!\! >
= (g_{j{\rm A}} m_{j{\rm A}} 
+ g_{j{\rm B}} m_{j{\rm B}}) \mu_B B
\end{equation}
where $g_j$ is given by the standard formula 
($g_j = 4/3$ for the $^{2}p_{3/2}$ states,
and $g_j = 6/5$ for the $^{2}d_{5/2}$ states \cite{kuhn:1962}).
In this basis set, with the axis of quantization along the direction
of the magnetic field, there are no off-diagonal elements of
$\hat{V}_{Z}$.

Once a magnetic field is applied only pairs of states for which
$m_{j,{\rm A}1}=m_{j,{\rm B}2}$  and
$m_{j,{\rm B}1}=m_{j,{\rm A}2}$,
are degenerate.  As an example, consider the state
$\!\!|46d_{5/2}m_{j}=1/2\!\!>_{\rm A}\!\!
|47p_{3/2}m_{j}=1/2\!\!>_{\rm B}$.
This is coupled by $\hat{V}_{dd}$ to several states, as shown
in Table~\ref{tb:example}.
These states are all energy degenerate in the absence of a
magnetic field and interactions.  
Once a magnetic  field is applied only one coupled
state has an identical Zeeman shift and remains degenerate:
$\!\!|47p_{3/2}m_{j}=1/2\!\!>_{\rm A}\!\!
|46d_{5/2}m_{j}=1/2\!\!>_{\rm B}$.
Since the application of a magnetic field lowers the energy 
degeneracy, we expect less line broadening since some of the
dipole-dipole couplings are no longer resonant.

\begin{table}
\caption{
\label{tb:example}
Non-zero matrix elements $<\!\!2|\hat{V}_{dd}|1\!\!>$ where
$|1\!\!\!>=\!\!|46d_{5/2}m_{j}\!\!=\!\!1/2\!\!>_{\rm A}\!\!
|47p_{3/2}m_{j}\!\!=\!\!1/2\!\!>_{\rm B}$ and
$|2\!\!\!>=\!\!|47p_{3/2}m_{j,{\rm A}}\!\!>_{\rm A}\!\!
|46d_{5/2}m_{j,{\rm B}}\!\!>_{\rm B}$, and 
$R_{\rm AB} = 28.5 \: {\rm \mu m}$.  
The axis for angular momentum quantization is in the direction
of $\vec{B}$ (see text). 
The dipole moments $\vec{\mu}_{\rm A}$ and $\vec{\mu}_{\rm B}$
in Eq.~\ref{eq:vdd} are evaluated using the techniques described
in Ref.~\cite{zimmerman:1979}.
The Zeeman 
energy shifts of the $|2\!\!>$ states relative to the
$|1\!\!>$ state are: 
{\( \mu_{\rm mag,2}-\mu_{\rm mag,1}
= [<\!\!2|\hat{V}_{Z}|2\!\!>-<\!\!1|\hat{V}_{Z}|1\!\!>]/B \)}
(calculated using Eq.~\ref{eq:zeeman}).
}

\setlength{\extrarowheight}{0.05in}
\begin{tabular}{r|r|c|D{.}{.}{6.0}|D{.}{.}{3.0}|D{.}{.}{4.0}} \hline\hline
\multicolumn{3}{c|}{$|2>$}
&
\multicolumn{2}{c|}{$<\!\!2 | \hat{V}_{dd} |1\!\!> \:$ (kHz)} &
\multicolumn{1}{c}{
$\mu_{\rm mag,2}-\mu_{\rm mag,1}$} \\ \cline{1-5}
$m_{j,{\rm A}}$ & $m_{j,{\rm B}}$ & $m_{j,{\rm tot}}$
& \multicolumn{1}{c|}{$\vec{n} \parallel \vec{B}$} & 
\multicolumn{1}{c|}{$\vec{n} \perp \vec{B}$} & 
\multicolumn{1}{c}{(kHz/G)} 
\\ 
\hline
$-1/2$ & $-1/2$ & $-1$ & 0 & 55 & -3547 \\
$3/2$ & $-1/2$ & $1$ & 21 & -11 & 187 \\
$1/2$ & $1/2$ & $1$ & -147 & 74 & 0 \\
$-1/2$ & $3/2$ & $1$ & 52 & -26 & -187 \\
$3/2$ & $3/2$ & $2$ & 0 & 45 & 3547 \\
\hline\hline
\end{tabular}
\setlength{\extrarowheight}{0.00in}
\end{table}

A quantitative treatment of the resulting linewidth
suppression will now be considered.
The essence of this treatment is the numerical diagonalization
of a Hamiltonian using a basis set of two atom states -- the
$24+24=48$ states mentioned above.
All of these 48 states are energy degenerate in the absence of
a magnetic field and dipole-dipole interactions, so the simplified
Hamiltonian is 
\begin{equation}
\hat{H} = \hat{V}_{Z} + \hat{V}_{dd}. 
\end{equation}
This corresponds to a 48 by 48 matrix for fixed values of
$\vec{n}$, $R_{\rm AB}$ and $B$.  
The matrix elements of $\hat{V}_{dd}$ require the transition dipole
moments $\vec{\mu}_{\rm A}$ and $\vec{\mu}_{\rm B}$ (see Eq.~\ref{eq:vdd}), 
which can be  evaluated using numerical integration of the Rydberg 
electron wave-functions \cite{zimmerman:1979}.  The matrix elements
of $\hat{V}_{Z}$ are obtained from Eq.~\ref{eq:zeeman}.
The resulting Hamiltonian matrix may be numerically diagonalized,
yielding a set of energy eigenvalues $E_{i}$ and corresponding
eigenvectors $|\psi_i \!\! >$.

As has been discussed, the two-photon probe transition is
not sensitive to the Zeeman effect \cite{li:2003}.  
Thus for a correspondence with the spectral transition energies, the
Zeeman energy contribution is subtracted from each eigenvalue:
\begin{equation}
E_{ci}=E_{i}-<\!\! \psi_i | \hat{V}_{Z}| \psi_i \!\!>.
\end{equation}
This heuristic expression is based on the observation that 
the probe linewidth is independent of magnetic field in the absence
of interatomic effects.  An exact treatment would take into account
relative linestrengths in the two-photon probe transition.

\begin{figure}
\includegraphics{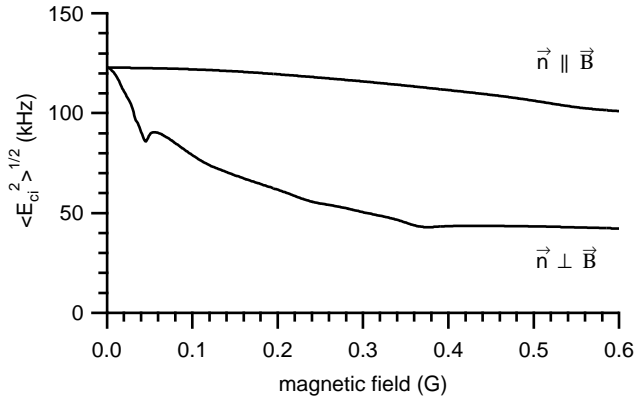}
\caption{\label{fg:soletheory}
The calculated RMS of the corrected energy 
eigenvalues $E_{ci}$ due to
the electric dipole-dipole interaction between 
$46d_{5/2}$ and $47p_{3/2}$ atoms, as a function
of magnetic field strength (see text).  The interatomic separation is
$R_{\rm AB} = 28.5 \: {\rm \mu m}$, 
the most probable resonantly interacting
neighbor distance at a total Rydberg density of 
$1.4 \times 10^7 \: {\rm cm}^{-3}$.
Calculations are shown for two different relative orientations of
the magnetic field $\vec{B}$, and the unit vector pointing between
the two atoms $\vec{n}$. 
}
\end{figure}

The corrected energies $E_{ci}$ may be computed as a
function of magnetic field magnitude
for a fixed $R_{\rm AB}$ and relative
orientation of $\vec{n}$ and $\vec{B}$.
The distribution of the corrected energies dictates the linewidth
broadening.  If there was no dipole-dipole interaction these
would all be zero.   The spread in the corrected energies decreases 
with magnetic
field -- the magnetic field suppresses the dipole-dipole interaction.
To quantify this with a single number, we have computed
the root mean square (RMS) corrected energy $<\!\!E_{ci}^2\!\!>^{1/2}$
as a function of magnetic field for $\vec{n} \perp \vec{B}$
and $\vec{n} \parallel \vec{B}$ at a fixed value of $R_{\rm AB}$
(see Fig. \ref{fg:soletheory}).

The qualititative difference in the suppression for 
{$\vec{n} \perp \vec{B}$} and $\vec{n} \parallel \vec{B}$ may
be understood from the difference in the selection rules
for $\hat{V}_{dd}$ in these two cases.
When $\vec{n}$ is parallel to the quantization axis ($\vec{B}$),
the dipole-dipole interaction $\hat{V}_{dd}$ does not couple
states of different $m_{j,{\rm tot}}$, where 
$m_{j,{\rm tot}} = m_{j,{\rm A}} + m_{j,{\rm B}}$.
This is a consequence of the invariance of $\hat{V}_{dd}$
to rotations about $\vec{n}$.  However the same selection rule
does not exist when $\vec{n}$ is perpendicular to the quantization
axis (see, for example, Table~\ref{tb:example}).  Due to the similarity
in $g_{j}$ factors for the $^{2}p_{3/2}$ 
and $^{2}d_{5/2}$ states ($4/3$ {\it vs.} $6/5$)
the Zeeman shifts for states of the same $m_{j,{\rm tot}}$
are quite similar when compared to states of different
$m_{j,{\rm tot}}$ (see the last column of Table~\ref{tb:example}).
Therefore the degeneracy between states coupled by
$\hat{V}_{dd}$ with $\vec{n} \parallel \vec{B}$ is spoiled
at much higher fields than for {$\vec{n} \perp \vec{B}$}.
If Fig.~\ref{fg:soletheory} is extended to higher fields,
then $<\!\!E_{ci}^2\!\!>^{1/2}$ for $\vec{n} \parallel \vec{B}$ 
eventually drops to 67 kHz.

In the sample both $\vec{n}$ and $R_{\rm AB}$ are distributed
randomly.  The nearest neighbor probability 
distribution is given by \cite{chandrasekhar:1943}:
\begin{equation}
\frac{dP(R_{\rm AB})}{dR_{\rm AB}} = 4\pi R_{\rm AB}^2 n_{\rm int}
\: \exp (-4\pi R_{\rm AB}^3 n_{\rm int} /3).
\end{equation}
The number density $n_{\rm int}$ 
in this expression is half of the total Rydberg density
-- as half of the atoms resonantly interact with the other half
in an even mixture of $46d_{5/2}$ and $47p_{3/2}$ atoms.
Using this expression
and random orientations of $\vec{n}$, 
we have calculated the probability 
distribution of the corrected eigenvalues.  
These distributions were convolved with a
${\rm sinc}^2(\pi f T)$ lineshape ($T=6 \: \mu s$)
and fitted in the same manner as the experimental data
(using the ${\rm sinc}^2(\pi f T)$ lineshape convolved with a 
Lorentzian of adjustable width $\delta\nu$).

To make a comparison with the experimental observations, the
calculations must be done at a specific number density.
As discussed, our experimental density determination is
uncertain by almost a factor of 2.  Therefore the density used
in the calculations is adjusted for the best fit to the data.
This is the only adjustable parameter in the calculations,
and we use the same adjustment factor for both densities in
Fig.~\ref{fg:suppression}.
Specifically, at the higher density the 
calculations are performed at a constant Rydberg density 
of $1.4 \times 10^7 \: {\rm cm^{-3}}$,
whereas the spatially averaged experimental density is estimated to be 
$(0.8 \pm 0.5) \times 10^7 \: {\rm cm^{-3}}$.  The discrepency
may be an indication of the importance of multi-body interactions,
which are expected to enhance the interactions over a simple binary model.

The calculations shown in Fig.~\ref{fg:suppression}
predict a reduction in linewidths 
with magnetic field comparable to the observations.
However the agreement is not perfect.  In particular, the experimental
results appear to extrapolate to a lower linewidth at zero magnetic
field.  There are
several possible reasons for this.  Hyperfine structure has 
been neglected.  This is expected to be especially important
at low magnetic fields where the Zeeman shifts are smaller
than the hyperfine splitting.  In addition, stray electric fields
break degeneracies between states of different $|m_j|$, and this
is also expected to suppress linewidth.

The distribution
of energy eigenvalues of the two atom system has been used
to obtain linewidths
-- but linestrengths for the different transitions should be
accounted for. In addition, not all of the eigenstates will be equally
populated by the excitation scheme.  

Perhaps the most significant problem with the theoretical model is the
assumption of binary interactions: only pairs of atoms interact.
Experimental observations of resonant energy transfer between
cold Rydberg atoms have demonstrated that this picture is not 
complete \cite{anderson:1998,mourachko:2004}.  Taking these into account
would require a much more 
sophisticated approach \cite{frasier:1999,robicheaux:2004}.

\vspace{0.1in}

\section{Concluding Remarks}

The magnetic field induced partial suppression of the 
resonant electric dipole-dipole
interaction between cold Rydberg atoms has been observed, and a quantitative
model of this phenomena has been presented.  This model could be
tested in more detail.  
A ``one-dimensional sample'' has recently
been successfully employed to determine the influence of the
relative orientation of an electric field with $\vec{n}$ on
Rydberg dipole-dipole interactions \cite{carroll:2004}.  
By focusing  the Rydberg state excitation
laser more tightly, so that the average interatomic separation
is larger than the beam waist, the sample becomes one dimensional
\cite{carroll:2004}.  
This gives a preferred direction to 
$\vec{n}$ in the lab-frame, and thus makes it possible to study 
interactions with $\vec{B} \perp \vec{n}$ or $\vec{B} \parallel \vec{n}$.
As Fig.~\ref{fg:soletheory} indicates, there should be a dramatic
difference between these two cases, which would provide a useful check.

Homogeneous DC magnetic fields cannot completely remove the energy
degeneracies responsible for the resonant dipole-dipole interaction.  
For example, in the absence of interactions 
$|46d_{5/2}m_{5/2}\!\!>_{\rm A}\!\! |47p_{3/2}m_{3/2}\!\!>_{\rm B}$ 
is energy degenerate with 
$|47p_{3/2}m_{3/2}\!\!>_{\rm A}\!\!|46d_{5/2}m_{5/2}\!\!>_{\rm B}$
independent of magnetic field.
For complete suppression of the dipole-dipole interaction it is
interesting to consider the use of magnetic field gradients.
By making the energy levels of the atoms depend on position,
the resonance conditions can be completely spoiled.  Strong magnetic
field gradients have recently been used in electron spin resonance
(ESR) to suppress spin
diffusion due to the magnetic dipolar flip-flop interaction 
\cite{budakian:2004}.
An analogous experiment could be performed in the case of
resonant electric dipole-dipole interactions.  The required field gradient
is modest.  For example, in the present work
we typically have 100 kHz interaction strengths at interatomic
separations of $30 \: {\rm \mu m}$.  With Zeeman shifts typically being
on the order of $1 \: {\rm MHz/G}$, field gradients of roughly
$50 \: {\rm G/cm}$ should give significant supression.  These gradients
are readily achievable -- in magnetic microtraps for example
\cite{folman:2002}.  Strong magnetic field gradients could be useful for
suppression of resonant dipole-dipole interactions in situations where 
they are undesirable.

It is a pleasure to acknowledge discussions with T.~F.~Gallagher.
We thank M.~Fedorov, D.~Vagale, and J.~C.~T.~Martin for 
assistance. This work was supported by NSERC, CFI, and OIT.  

\bibliography{references}

\begin{thebibliography}{34}
\expandafter\ifx\csname natexlab\endcsname\relax\def\natexlab#1{#1}\fi
\expandafter\ifx\csname bibnamefont\endcsname\relax
  \def\bibnamefont#1{#1}\fi
\expandafter\ifx\csname bibfnamefont\endcsname\relax
  \def\bibfnamefont#1{#1}\fi
\expandafter\ifx\csname citenamefont\endcsname\relax
  \def\citenamefont#1{#1}\fi
\expandafter\ifx\csname url\endcsname\relax
  \def\url#1{\texttt{#1}}\fi
\expandafter\ifx\csname urlprefix\endcsname\relax\def\urlprefix{URL }\fi
\providecommand{\bibinfo}[2]{#2}
\providecommand{\eprint}[2][]{\url{#2}}

\bibitem[{\citenamefont{Anderson et~al.}(1998)\citenamefont{Anderson, Veale,
  and Gallagher}}]{anderson:1998}
\bibinfo{author}{\bibfnamefont{W.~R.} \bibnamefont{Anderson}},
  \bibinfo{author}{\bibfnamefont{J.~R.} \bibnamefont{Veale}}, \bibnamefont{and}
  \bibinfo{author}{\bibfnamefont{T.~F.} \bibnamefont{Gallagher}},
  \bibinfo{journal}{Phys. Rev. Lett.} \textbf{\bibinfo{volume}{80}},
  \bibinfo{pages}{249} (\bibinfo{year}{1998}).

\bibitem[{\citenamefont{Mourachko et~al.}(1998)\citenamefont{Mourachko,
  Comparat, de~Tomasi, Fioretti, Nosbaum, Akulin, and Pillet}}]{mourachko:1998}
\bibinfo{author}{\bibfnamefont{I.}~\bibnamefont{Mourachko}},
  \bibinfo{author}{\bibfnamefont{D.}~\bibnamefont{Comparat}},
  \bibinfo{author}{\bibfnamefont{F.}~\bibnamefont{de~Tomasi}},
  \bibinfo{author}{\bibfnamefont{A.}~\bibnamefont{Fioretti}},
  \bibinfo{author}{\bibfnamefont{P.}~\bibnamefont{Nosbaum}},
  \bibinfo{author}{\bibfnamefont{V.~M.} \bibnamefont{Akulin}},
  \bibnamefont{and} \bibinfo{author}{\bibfnamefont{P.}~\bibnamefont{Pillet}},
  \bibinfo{journal}{Phys. Rev. Lett.} \textbf{\bibinfo{volume}{80}},
  \bibinfo{pages}{253} (\bibinfo{year}{1998}).

\bibitem[{\citenamefont{Li et~al.}(2004)}]{li:2004}
\bibinfo{author}{\bibfnamefont{W.}~\bibnamefont{Li}} \bibnamefont{et~al.},
  \bibinfo{journal}{Phys. Rev. A} \textbf{\bibinfo{volume}{70}},
  \bibinfo{pages}{042713} (\bibinfo{year}{2004}).

\bibitem[{\citenamefont{Robicheaux et~al.}(2004)\citenamefont{Robicheaux,
  Hern\'{a}ndez, Top\c{c}u, and Noordam}}]{robicheaux:2004}
\bibinfo{author}{\bibfnamefont{F.}~\bibnamefont{Robicheaux}},
  \bibinfo{author}{\bibfnamefont{J.~V.} \bibnamefont{Hern\'{a}ndez}},
  \bibinfo{author}{\bibfnamefont{T.}~\bibnamefont{Top\c{c}u}},
  \bibnamefont{and} \bibinfo{author}{\bibfnamefont{L.~D.}
  \bibnamefont{Noordam}}, \bibinfo{journal}{Phys. Rev. A}
  \textbf{\bibinfo{volume}{70}}, \bibinfo{pages}{042703}
  (\bibinfo{year}{2004}).

\bibitem[{\citenamefont{Jaksch et~al.}(2000)\citenamefont{Jaksch, Cirac,
  Zoller, Rolston, C\^{o}t\'{e}, and Lukin}}]{jaksch:2000}
\bibinfo{author}{\bibfnamefont{D.}~\bibnamefont{Jaksch}},
  \bibinfo{author}{\bibfnamefont{J.~I.} \bibnamefont{Cirac}},
  \bibinfo{author}{\bibfnamefont{P.}~\bibnamefont{Zoller}},
  \bibinfo{author}{\bibfnamefont{S.~L.} \bibnamefont{Rolston}},
  \bibinfo{author}{\bibfnamefont{R.}~\bibnamefont{C\^{o}t\'{e}}},
  \bibnamefont{and} \bibinfo{author}{\bibfnamefont{M.~D.} \bibnamefont{Lukin}},
  \bibinfo{journal}{Phys. Rev. Lett.} \textbf{\bibinfo{volume}{85}},
  \bibinfo{pages}{2208} (\bibinfo{year}{2000}).

\bibitem[{\citenamefont{Lukin et~al.}(2001)\citenamefont{Lukin, Fleischhauer,
  Cote, Duan, Jaksch, Cirac, and Zoller}}]{lukin:2001}
\bibinfo{author}{\bibfnamefont{M.~D.} \bibnamefont{Lukin}},
  \bibinfo{author}{\bibfnamefont{M.}~\bibnamefont{Fleischhauer}},
  \bibinfo{author}{\bibfnamefont{R.}~\bibnamefont{Cote}},
  \bibinfo{author}{\bibfnamefont{L.~M.} \bibnamefont{Duan}},
  \bibinfo{author}{\bibfnamefont{D.}~\bibnamefont{Jaksch}},
  \bibinfo{author}{\bibfnamefont{J.~I.} \bibnamefont{Cirac}}, \bibnamefont{and}
  \bibinfo{author}{\bibfnamefont{P.}~\bibnamefont{Zoller}},
  \bibinfo{journal}{Phys. Rev. Lett.} \textbf{\bibinfo{volume}{87}},
  \bibinfo{pages}{037901} (\bibinfo{year}{2001}).

\bibitem[{\citenamefont{Saffman and Walker}(2002)}]{saffman:2002}
\bibinfo{author}{\bibfnamefont{M.}~\bibnamefont{Saffman}} \bibnamefont{and}
  \bibinfo{author}{\bibfnamefont{T.~G.} \bibnamefont{Walker}},
  \bibinfo{journal}{Phys. Rev. A} \textbf{\bibinfo{volume}{66}},
  \bibinfo{pages}{065403} (\bibinfo{year}{2002}).

\bibitem[{\citenamefont{Tong et~al.}(2004)\citenamefont{Tong, Farooqi,
  Stanojevic, Krishnan, Zhang, C\^{o}t\'{e}, Eyler, and Gould}}]{tong:2004}
\bibinfo{author}{\bibfnamefont{D.}~\bibnamefont{Tong}},
  \bibinfo{author}{\bibfnamefont{S.~M.} \bibnamefont{Farooqi}},
  \bibinfo{author}{\bibfnamefont{J.}~\bibnamefont{Stanojevic}},
  \bibinfo{author}{\bibfnamefont{S.}~\bibnamefont{Krishnan}},
  \bibinfo{author}{\bibfnamefont{Y.~P.} \bibnamefont{Zhang}},
  \bibinfo{author}{\bibfnamefont{R.}~\bibnamefont{C\^{o}t\'{e}}},
  \bibinfo{author}{\bibfnamefont{E.~E.} \bibnamefont{Eyler}}, \bibnamefont{and}
  \bibinfo{author}{\bibfnamefont{P.~L.} \bibnamefont{Gould}},
  \bibinfo{journal}{Phys. Rev. Lett.} \textbf{\bibinfo{volume}{93}},
  \bibinfo{pages}{063001} (\bibinfo{year}{2004}).

\bibitem[{\citenamefont{Singer et~al.}(2004)\citenamefont{Singer, Reetz-Lamour,
  Amthor, Marcassa, and Weidem\"{u}ller}}]{singer:2004}
\bibinfo{author}{\bibfnamefont{K.}~\bibnamefont{Singer}},
  \bibinfo{author}{\bibfnamefont{M.}~\bibnamefont{Reetz-Lamour}},
  \bibinfo{author}{\bibfnamefont{T.}~\bibnamefont{Amthor}},
  \bibinfo{author}{\bibfnamefont{L.~G.} \bibnamefont{Marcassa}},
  \bibnamefont{and}
  \bibinfo{author}{\bibfnamefont{M.}~\bibnamefont{Weidem\"{u}ller}},
  \bibinfo{journal}{Phys. Rev. Lett.} \textbf{\bibinfo{volume}{93}},
  \bibinfo{pages}{163001} (\bibinfo{year}{2004}).

\bibitem[{\citenamefont{Ryabtsev et~al.}(2005)\citenamefont{Ryabtsev,
  Tretyakov, and Beterov}}]{ryabtsev:2005}
\bibinfo{author}{\bibfnamefont{I.~I.} \bibnamefont{Ryabtsev}},
  \bibinfo{author}{\bibfnamefont{D.~B.} \bibnamefont{Tretyakov}},
  \bibnamefont{and} \bibinfo{author}{\bibfnamefont{I.~I.}
  \bibnamefont{Beterov}}, \bibinfo{journal}{J. Phys. B: At. Mol. Opt. Phys.}
  \textbf{\bibinfo{volume}{38}}, \bibinfo{pages}{S421} (\bibinfo{year}{2005}).

\bibitem[{\citenamefont{Mourachko et~al.}(2004)\citenamefont{Mourachko, Li, and
  Gallagher}}]{mourachko:2004}
\bibinfo{author}{\bibfnamefont{I.}~\bibnamefont{Mourachko}},
  \bibinfo{author}{\bibfnamefont{W.}~\bibnamefont{Li}}, \bibnamefont{and}
  \bibinfo{author}{\bibfnamefont{T.~F.} \bibnamefont{Gallagher}},
  \bibinfo{journal}{Phys. Rev. A} \textbf{\bibinfo{volume}{70}},
  \bibinfo{pages}{031401(R)} (\bibinfo{year}{2004}).

\bibitem[{\citenamefont{Li et~al.}(2005)\citenamefont{Li, Tanner, and
  Gallagher}}]{li:2005}
\bibinfo{author}{\bibfnamefont{W.}~\bibnamefont{Li}},
  \bibinfo{author}{\bibfnamefont{P.~J.} \bibnamefont{Tanner}},
  \bibnamefont{and} \bibinfo{author}{\bibfnamefont{T.~F.}
  \bibnamefont{Gallagher}}, \bibinfo{journal}{Phys. Rev. Lett.}
  \textbf{\bibinfo{volume}{94}}, \bibinfo{pages}{173001}
  (\bibinfo{year}{2005}).

\bibitem[{\citenamefont{Afrousheh et~al.}(2004)\citenamefont{Afrousheh,
  Bohlouli-Zanjani, Vagale, Mugford, Fedorov, and Martin}}]{afrousheh:2004}
\bibinfo{author}{\bibfnamefont{K.}~\bibnamefont{Afrousheh}},
  \bibinfo{author}{\bibfnamefont{P.}~\bibnamefont{Bohlouli-Zanjani}},
  \bibinfo{author}{\bibfnamefont{D.}~\bibnamefont{Vagale}},
  \bibinfo{author}{\bibfnamefont{A.}~\bibnamefont{Mugford}},
  \bibinfo{author}{\bibfnamefont{M.}~\bibnamefont{Fedorov}}, \bibnamefont{and}
  \bibinfo{author}{\bibfnamefont{J.~D.~D.} \bibnamefont{Martin}},
  \bibinfo{journal}{Phys. Rev. Lett.} \textbf{\bibinfo{volume}{93}},
  \bibinfo{pages}{233001} (\bibinfo{year}{2004}).

\bibitem[{\citenamefont{Lesanovsky and Schmelcher}(2005)}]{lesanovsky:2005}
\bibinfo{author}{\bibfnamefont{I.}~\bibnamefont{Lesanovsky}} \bibnamefont{and}
  \bibinfo{author}{\bibfnamefont{P.}~\bibnamefont{Schmelcher}},
  \bibinfo{journal}{Phys. Rev. Lett.} \textbf{\bibinfo{volume}{95}},
  \bibinfo{pages}{053001} (\bibinfo{year}{2005}).

\bibitem[{\citenamefont{Monroe et~al.}(1990)\citenamefont{Monroe, Swann,
  Robinson, and Wieman}}]{monroe:1990}
\bibinfo{author}{\bibfnamefont{C.}~\bibnamefont{Monroe}},
  \bibinfo{author}{\bibfnamefont{W.}~\bibnamefont{Swann}},
  \bibinfo{author}{\bibfnamefont{H.}~\bibnamefont{Robinson}}, \bibnamefont{and}
  \bibinfo{author}{\bibfnamefont{C.}~\bibnamefont{Wieman}},
  \bibinfo{journal}{Phys. Rev. Lett.} \textbf{\bibinfo{volume}{65}},
  \bibinfo{pages}{1571} (\bibinfo{year}{1990}).

\bibitem[{\citenamefont{Gallagher}(1994)}]{gallagher:1994}
\bibinfo{author}{\bibfnamefont{T.~F.} \bibnamefont{Gallagher}},
  \emph{\bibinfo{title}{Rydberg Atoms}} (\bibinfo{publisher}{Cambridge
  University Press}, \bibinfo{year}{1994}).

\bibitem[{\citenamefont{Teo et~al.}(2003)\citenamefont{Teo, Feldbaum, Cubel,
  Guest, Berman, and Raithel}}]{teo:2003}
\bibinfo{author}{\bibfnamefont{B.~K.} \bibnamefont{Teo}},
  \bibinfo{author}{\bibfnamefont{D.}~\bibnamefont{Feldbaum}},
  \bibinfo{author}{\bibfnamefont{T.}~\bibnamefont{Cubel}},
  \bibinfo{author}{\bibfnamefont{J.~R.} \bibnamefont{Guest}},
  \bibinfo{author}{\bibfnamefont{P.~R.} \bibnamefont{Berman}},
  \bibnamefont{and} \bibinfo{author}{\bibfnamefont{G.}~\bibnamefont{Raithel}},
  \bibinfo{journal}{Phys. Rev. A} \textbf{\bibinfo{volume}{68}},
  \bibinfo{pages}{053407} (\bibinfo{year}{2003}).

\bibitem[{\citenamefont{Osterwalder and Merkt}(1999)}]{osterwalder:1999}
\bibinfo{author}{\bibfnamefont{A.}~\bibnamefont{Osterwalder}} \bibnamefont{and}
  \bibinfo{author}{\bibfnamefont{F.}~\bibnamefont{Merkt}},
  \bibinfo{journal}{Phys. Rev. Lett.} \textbf{\bibinfo{volume}{82}},
  \bibinfo{pages}{1831} (\bibinfo{year}{1999}).

\bibitem[{\citenamefont{Suzaki and Tachibana}(1975)}]{suzaki:1975}
\bibinfo{author}{\bibfnamefont{Y.}~\bibnamefont{Suzaki}} \bibnamefont{and}
  \bibinfo{author}{\bibfnamefont{A.}~\bibnamefont{Tachibana}},
  \bibinfo{journal}{Appl. Opt.} \textbf{\bibinfo{volume}{14}},
  \bibinfo{pages}{2809} (\bibinfo{year}{1975}).

\bibitem[{\citenamefont{Fraser}(2002)}]{fraser:2002}
\bibinfo{author}{\bibfnamefont{G.~W.} \bibnamefont{Fraser}},
  \bibinfo{journal}{Int. J. Mass Spectrom.} \textbf{\bibinfo{volume}{215}},
  \bibinfo{pages}{13} (\bibinfo{year}{2002}).

\bibitem[{\citenamefont{Li et~al.}(2003)\citenamefont{Li, Mourachko, Noel, and
  Gallagher}}]{li:2003}
\bibinfo{author}{\bibfnamefont{W.}~\bibnamefont{Li}},
  \bibinfo{author}{\bibfnamefont{I.}~\bibnamefont{Mourachko}},
  \bibinfo{author}{\bibfnamefont{M.~W.} \bibnamefont{Noel}}, \bibnamefont{and}
  \bibinfo{author}{\bibfnamefont{T.~F.} \bibnamefont{Gallagher}},
  \bibinfo{journal}{Phys. Rev. A} \textbf{\bibinfo{volume}{67}},
  \bibinfo{pages}{052502} (\bibinfo{year}{2003}).

\bibitem[{\citenamefont{Zimmerman et~al.}(1979)\citenamefont{Zimmerman,
  Littman, Kash, and Kleppner}}]{zimmerman:1979}
\bibinfo{author}{\bibfnamefont{M.~L.} \bibnamefont{Zimmerman}},
  \bibinfo{author}{\bibfnamefont{M.~G.} \bibnamefont{Littman}},
  \bibinfo{author}{\bibfnamefont{M.~M.} \bibnamefont{Kash}}, \bibnamefont{and}
  \bibinfo{author}{\bibfnamefont{D.}~\bibnamefont{Kleppner}},
  \bibinfo{journal}{Phys. Rev. A} \textbf{\bibinfo{volume}{20}},
  \bibinfo{pages}{2251} (\bibinfo{year}{1979}).

\bibitem[{\citenamefont{Fioretti et~al.}(1999)\citenamefont{Fioretti, Comparat,
  Drag, Gallagher, and Pillet}}]{fioretti:1999}
\bibinfo{author}{\bibfnamefont{A.}~\bibnamefont{Fioretti}},
  \bibinfo{author}{\bibfnamefont{D.}~\bibnamefont{Comparat}},
  \bibinfo{author}{\bibfnamefont{C.}~\bibnamefont{Drag}},
  \bibinfo{author}{\bibfnamefont{T.~F.} \bibnamefont{Gallagher}},
  \bibnamefont{and} \bibinfo{author}{\bibfnamefont{P.}~\bibnamefont{Pillet}},
  \bibinfo{journal}{Phys. Rev. Lett.} \textbf{\bibinfo{volume}{82}},
  \bibinfo{pages}{1839} (\bibinfo{year}{1999}).

\bibitem[{\citenamefont{Anderson et~al.}(2002)\citenamefont{Anderson, Robinson,
  Martin, and Gallagher}}]{anderson:2002}
\bibinfo{author}{\bibfnamefont{W.~R.} \bibnamefont{Anderson}},
  \bibinfo{author}{\bibfnamefont{M.~P.} \bibnamefont{Robinson}},
  \bibinfo{author}{\bibfnamefont{J.~D.~D.} \bibnamefont{Martin}},
  \bibnamefont{and} \bibinfo{author}{\bibfnamefont{T.~F.}
  \bibnamefont{Gallagher}}, \bibinfo{journal}{Phys. Rev. A}
  \textbf{\bibinfo{volume}{65}}, \bibinfo{pages}{063404}
  (\bibinfo{year}{2002}).

\bibitem[{\citenamefont{Robinson et~al.}(2000)\citenamefont{Robinson, Tolra,
  Noel, Gallagher, and Pillet}}]{robinson:2000}
\bibinfo{author}{\bibfnamefont{M.~P.} \bibnamefont{Robinson}},
  \bibinfo{author}{\bibfnamefont{B.~L.} \bibnamefont{Tolra}},
  \bibinfo{author}{\bibfnamefont{M.~W.} \bibnamefont{Noel}},
  \bibinfo{author}{\bibfnamefont{T.~F.} \bibnamefont{Gallagher}},
  \bibnamefont{and} \bibinfo{author}{\bibfnamefont{P.}~\bibnamefont{Pillet}},
  \bibinfo{journal}{Phys. Rev. Lett.} \textbf{\bibinfo{volume}{85}},
  \bibinfo{pages}{4466} (\bibinfo{year}{2000}).

\bibitem[{\citenamefont{Fortagh et~al.}(1998)\citenamefont{Fortagh, Grossmann,
  H\"{a}nsch, and Zimmermann}}]{fortagh:1998}
\bibinfo{author}{\bibfnamefont{J.}~\bibnamefont{Fortagh}},
  \bibinfo{author}{\bibfnamefont{A.}~\bibnamefont{Grossmann}},
  \bibinfo{author}{\bibfnamefont{T.~W.} \bibnamefont{H\"{a}nsch}},
  \bibnamefont{and}
  \bibinfo{author}{\bibfnamefont{C.}~\bibnamefont{Zimmermann}},
  \bibinfo{journal}{J. Appl. Phys.} \textbf{\bibinfo{volume}{84}},
  \bibinfo{pages}{6499} (\bibinfo{year}{1998}).

\bibitem[{\citenamefont{Dedman et~al.}(2001)\citenamefont{Dedman, Baldwin, and
  Colla}}]{dedman:2001}
\bibinfo{author}{\bibfnamefont{C.~J.} \bibnamefont{Dedman}},
  \bibinfo{author}{\bibfnamefont{K.~G.~H.} \bibnamefont{Baldwin}},
  \bibnamefont{and} \bibinfo{author}{\bibfnamefont{M.}~\bibnamefont{Colla}},
  \bibinfo{journal}{Rev. Sci. Instrum.} \textbf{\bibinfo{volume}{72}},
  \bibinfo{pages}{4055} (\bibinfo{year}{2001}).

\bibitem[{\citenamefont{Kuhn}(1962)}]{kuhn:1962}
\bibinfo{author}{\bibfnamefont{H.~G.} \bibnamefont{Kuhn}},
  \emph{\bibinfo{title}{Atomic Spectra}} (\bibinfo{publisher}{Longmans},
  \bibinfo{address}{London}, \bibinfo{year}{1962}).

\bibitem[{\citenamefont{Santra and Greene}(2003)}]{santra:2003}
\bibinfo{author}{\bibfnamefont{R.}~\bibnamefont{Santra}} \bibnamefont{and}
  \bibinfo{author}{\bibfnamefont{C.~H.} \bibnamefont{Greene}},
  \bibinfo{journal}{Phys. Rev. A} \textbf{\bibinfo{volume}{67}},
  \bibinfo{pages}{062713} (\bibinfo{year}{2003}).

\bibitem[{\citenamefont{Chandrasekhar}(1943)}]{chandrasekhar:1943}
\bibinfo{author}{\bibfnamefont{S.}~\bibnamefont{Chandrasekhar}},
  \bibinfo{journal}{Rev. Mod. Phys.} \textbf{\bibinfo{volume}{15}},
  \bibinfo{pages}{1} (\bibinfo{year}{1943}).

\bibitem[{\citenamefont{Frasier et~al.}(1999)\citenamefont{Frasier, Celli, and
  Blum}}]{frasier:1999}
\bibinfo{author}{\bibfnamefont{J.~S.} \bibnamefont{Frasier}},
  \bibinfo{author}{\bibfnamefont{V.}~\bibnamefont{Celli}}, \bibnamefont{and}
  \bibinfo{author}{\bibfnamefont{T.}~\bibnamefont{Blum}},
  \bibinfo{journal}{Phys. Rev. A} \textbf{\bibinfo{volume}{59}},
  \bibinfo{pages}{4358} (\bibinfo{year}{1999}).

\bibitem[{\citenamefont{Carroll et~al.}(2004)\citenamefont{Carroll,
  Claringbould, Goodsell, Lim, and Noel}}]{carroll:2004}
\bibinfo{author}{\bibfnamefont{T.~J.} \bibnamefont{Carroll}},
  \bibinfo{author}{\bibfnamefont{K.}~\bibnamefont{Claringbould}},
  \bibinfo{author}{\bibfnamefont{A.}~\bibnamefont{Goodsell}},
  \bibinfo{author}{\bibfnamefont{M.~J.} \bibnamefont{Lim}}, \bibnamefont{and}
  \bibinfo{author}{\bibfnamefont{M.~W.} \bibnamefont{Noel}},
  \bibinfo{journal}{Phys. Rev. Lett.} \textbf{\bibinfo{volume}{93}},
  \bibinfo{pages}{153001} (\bibinfo{year}{2004}).

\bibitem[{\citenamefont{Budakian et~al.}(2004)\citenamefont{Budakian, Mamin,
  and Rugar}}]{budakian:2004}
\bibinfo{author}{\bibfnamefont{R.}~\bibnamefont{Budakian}},
  \bibinfo{author}{\bibfnamefont{H.~J.} \bibnamefont{Mamin}}, \bibnamefont{and}
  \bibinfo{author}{\bibfnamefont{D.}~\bibnamefont{Rugar}},
  \bibinfo{journal}{Phys. Rev. Lett.} \textbf{\bibinfo{volume}{92}},
  \bibinfo{pages}{037205} (\bibinfo{year}{2004}).

\bibitem[{\citenamefont{Folman et~al.}(2002)\citenamefont{Folman, Kr\"{u}ger,
  Schmiedmayer, Denschlag, and Henkel}}]{folman:2002}
\bibinfo{author}{\bibfnamefont{R.}~\bibnamefont{Folman}},
  \bibinfo{author}{\bibfnamefont{P.}~\bibnamefont{Kr\"{u}ger}},
  \bibinfo{author}{\bibfnamefont{J.}~\bibnamefont{Schmiedmayer}},
  \bibinfo{author}{\bibfnamefont{J.}~\bibnamefont{Denschlag}},
  \bibnamefont{and} \bibinfo{author}{\bibfnamefont{C.}~\bibnamefont{Henkel}},
  \bibinfo{journal}{Adv. At. Mol. Opt. Phys.} \textbf{\bibinfo{volume}{48}},
  \bibinfo{pages}{263} (\bibinfo{year}{2002}).

\end{thebibliography}

\end{document}